\documentclass[aps,prl,preprint,groupedaddress,showpacs]{revtex4}

\usepackage{graphicx}
\usepackage{bm}

\begin{document}

\title{Measurement of the electron electric dipole moment using GdIG}

\author{B. J. Heidenreich}
\author{O. T. Elliott}
\author{N. D. Charney}
\author{K. A. Virgien}
\author{A. W. Bridges}
\author{M. A. McKeon}
\author{S. K. Peck}
\author{D. \surname{Krause}, Jr.}
\author{J. E. Gordon}
\author{L. R. Hunter}
\affiliation{Physics Department, Amherst College, Amherst, MA 01002}
\author{S. K. Lamoreaux}
\affiliation{Los Alamos National Laboratory, Physics Division, Los Alamos, NM 87545}

\date{September 12, 2005}

\begin{abstract}
A new method for the detection of the electron edm using a solid is described.  The method involves the measurement of a voltage induced across the solid by the alignment of the sample's magnetic dipoles in an applied magnetic field, $H$.  A first application of the method to GdIG has resulted in a limit on the electron edm of $5\times10^{-24}$~\mbox{e-cm}, which is a factor of 40 below the limit obtained from the only previous solid-state edm experiment. The result is limited by the imperfect discrimination of an unexpectedly large voltage that is even upon the reversal of the sample magnetization.
\end{abstract}


\maketitle
Precision measurement of the electron edm provides one of the most important low energy tests of time-reversal symmetry (T) and of particle physics beyond the standard model (SM)~\cite{commins99}. T violation in the SM is insufficient to explain the observed baryon asymmetry in the universe, suggesting new T-violating physics beyond the SM~\cite{balazs05}. The present upper bound on the electron edm, $d_e < 1.6 \times10^{-27}$~\mbox{e-cm}, already seriously constrains supersymmetry and many other SM extensions~\cite{regan02}. Several new experimental approaches are now being pursued that hope to achieve sensitivities to the electron edm beyond those established by the traditional heavy atom experiments in beams and cells~\cite{murthy89,peck94}. Cold trapped atoms~\cite{weiss03,heinzen}, diatomic molecules~\cite{demille00,hudson02}, trapped molecular ions~\cite{cornell}, and solid state systems~\cite{lamoreaux02} are all being investigated.

The idea to search for the electron edm in a solid was first suggested by Shapiro~\cite{shapiro68}. All proposed solid-state methods rely on the fact that the electron edm must be collinear with its magnetic moment, thus producing a linear magneto-electric (M-E) effect.  The only previous solid-state edm experiment, done using nickel-zinc ferrite, yielded a limit on the electron edm of $d_e < 2\times10^{-22}$~\mbox{e-cm}~\cite{vasilev78}. In this experiment SQUIDs were used to search for a change in the sample magnetization with the reversal of an applied electric field.  Presently, a refinement of this approach using gadolinium-gallium garnet (GdGG) is being pursued at Los Alamos.  Our approach is the compliment of this method.  We look for a change in the voltage across a solid state sample when the magnetization of the sample is reversed.

We perform our search in gadolinium-iron-garnet (Gd$_3^{3+}$Fe$_2^{3+}$Fe$_3^{3+}$O$_{12}$).  The sensitivity of GdIG to the electron edm comes through the Gd$^{3+}$ ions.  These ions have a high electronic spin (7/2), no orbital angular momentum, and a favorable heavy-ion enhancement factor~\cite{dzuba02}. In the absence of T violation, the crystal symmetry of the GdIG forbids terms in the free energy that go like $HE$ (the linear M-E effect) as well as terms of the form $H^2E$, where $H$ and $E$ are the magnetic and electric fields~\cite{mercier74}. The iron-garnets have high electrical resistivity that limits the migration of free charges that might otherwise cancel the edm induced dipole in the sample.  The presence of the ferromagnetic iron lattices allows the sample to be magnetized at relatively modest applied fields and temperatures.  Calculations have been undertaken that suggest that the existence of an electron edm with a magnitude of the present upper bound would produce 1.1~nV across a pure GdIG sample 10~cm in length and at 0~K~\cite{mukhamedjanov03}. A well designed detector would permit the measurement of such a voltage in a few hundred seconds, suggesting that with realistic integration times one might achieve a statistical limit on the electron edm one to two orders of magnitude better than the present bound.

\begin{figure}
\includegraphics{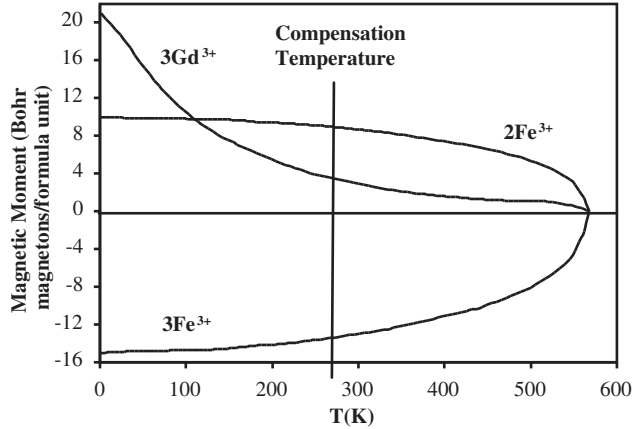}
\caption{Magnetization of the iron and gadolinium lattices in pure GdIG as a function of temperature. In the region above (below) the compensation temperature, the net magnetization is aligned antiparallel (parallel) to the Gd spin.\label{gdigmag}}
\end{figure}

GdIG is ferrimagnetic and three different lattices contribute to its magnetization (Fig.~\ref{gdigmag})~\cite{geller61}. At 0~K, two iron lattices produce a magnetization per unit cell that sums to 5~$\mu_B$ while the Gd$^{3+}$ ions produce a magnetization of 21~$\mu_B$ in the opposite direction.  While the magnetization of the paramagnetic Gd$^{3+}$ ions falls dramatically as the sample temperature is increased to room temperature, the iron magnetization falls off only slowly producing a compensation temperature ($T_C$) at 290~K.  At this temperature the sample has no net magnetization while above (below) the magnetization is dominated by the iron (Gd) lattice.  The contribution to the magnetization made by the Gd$^{3+}$ ions can be reduced by replacing some of them with non magnetic yttrium (Y).  The resulting compensation temperature can then be adjusted according to the formula: $T_C = (290 -115(3-x))$~K, where $x$ represents the average number of Gd and $3-x$ represents the average number of Y per unit cell~\cite{vonaulock65}.

\begin{figure}
\includegraphics{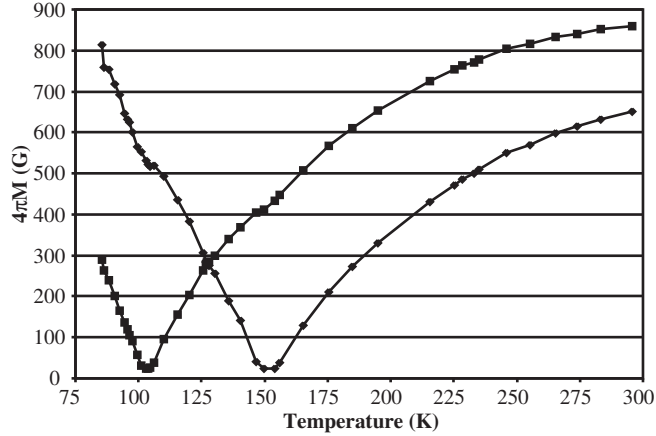}
\caption{Observed magnetization of the two ``C''s with an applied field of 40~Oe.  Squares (diamonds) correspond to the ``C'' with 1.35 (1.8) Gd per unit formula cell.\label{observedmag}}
\end{figure}

\begin{figure}
\includegraphics{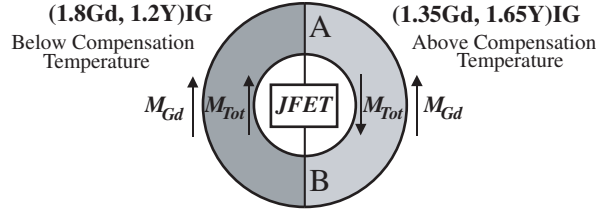}
\caption{Gd and total magnetizations of the toroid at the signal temperature of 127~K (topdown view.) The electrodes are designated by the letters $A$ and $B$ as shown.\label{toroid}}
\end{figure}

We take advantage of this interesting temperature and Gd concentration dependence to produce a distinctive temperature signature for the edm signal.  Our sample is a 2'' high toroid (4''~od and 2''~id) with rectangular cross section.  The toroidal geometry minimizes the role of demagnetizing fields, permitting the magnetization of the sample at modest applied fields.  The geometry also allows the placement of the detector in a nearly field free region.  The toroid is assembled from two ``C''s that are geometrically identical.  The electrodes are formed by bonding two 0.001'' thick copper foils between the ``C''s using conductive epoxy.  ``C1'' has $x = 1.35$ and an observed compensation temperature of $T_{C1} = 103$~K while ``C2'' has $x = 1.8$ and an observed $T_{C2} = 154$~K.  At the signal temperature, $T_S = 127$~K, the magnetizations ($M$) of the two ``C''s are identical (Fig.~\ref{observedmag}).  However, $T_{C1} < T_S$ and hence the net magnetization of C1 is antiparallel to the magnetization of its Gd$^{3+}$ ions while $T_{C2} > T_S$ and the net magnetization of C2 is parallel to the magnetization of its Gd$^{3+}$ ions.  Hence, when a toroidal magnetic field $\mathbf{H}$ is applied to the sample, the Gd$^{3+}$ ions are all oriented towards the same electrode (Fig.~\ref{toroid}) and the resulting edm signals from the two halves add constructively.  By contrast, at temperatures below 103~K (above 154~K), the Gd$^{3+}$ magnetization will be parallel (anti-parallel) to $\mathbf{M}$ on both sides of the toroid, resulting in a strong cancellation of the edm voltage between the two electrodes.  This dramatic change in the edm voltage with toroid temperature allows one to distinguish the edm voltage from voltage sources that might otherwise mimic the edm signal.

The sample is mounted in a toroid shaped copper Faraday cage (5.5'' high, 6''~od, 1.25''~id).  Heating elements attached directly to the outside of the copper cage enable temperature regulation.  A 336 turn coil of \#17 magnet wire is wound on the outside of the Faraday cage to produce $\mathbf{H}$.  Wires attached to the two electrodes at the outer radius of the sample ascend vertically through two thin-walled stainless steel tubes to the preamplifiers.  The preamplifier for each electrode consists of a Cascode pair of BF 245 JFETs and has an input impedance of $10^{13}$~$\Omega$ and an input capacitance of about 4~pF.  The preamplifiers are mounted in a copper and brass chamber with independent temperature monitoring and control.  The preamplifier, Faraday cage and sample are all contained within a vacuum system that can be immersed in liquid nitrogen.  The signals from the preamplifier are AC coupled (time constant of 5~s) out of the vacuum chamber and further amplified before being recorded on a digital oscilloscope.   

Data collection consists of averaging (typically between 128 and 512 traces) and recording the amplified electrode voltages as the direction of the sample magnetization is reversed (0.9~Hz -- 2.3~Hz).  We collect data in two different modes.  In the ``square wave'' mode, $\mathbf{H}$ is simply flipped from one orientation and back again.  $\mathbf{H}$ remains on during data acquisition.  In this mode we are usually limited by heating to applied fields of about 53~Oe, though for short periods of time we have operated with up to 124~Oe.  

In the ``pulsed'' mode, $\mathbf{H}$ is applied in a short pulse (typically between 7 and 50~ms) with amplitudes of up to 425~Oe.  Following a pulse, demagnetizing fields quickly result in the decay of the magnetization of the more magnetized half of the sample to the remnant magnetization of the less magnetized half.  With matched magnetizations the toroid has only an anapole moment and the remnant magnetization stabilizes and persists for many hours after the magnetizing pulse.  In this mode, the applied field is not present during data acquisition.  Normally, we collect edm data at three operating temperatures with approximately equal remnant magnetizations (Fig.~\ref{observedmag}):  88~K, 127~K, and 178~K.  In the pulsed mode, these operating temperatures have relative sensitivities to the electron edm of $-0.4$, 1, and $-0.3$ respectively, producing an easily discriminated temperature signature.

Induction pulses can be seen on the electrodes during the application and removal of $\mathbf{H}$.  These induction voltages reverse with the polarity of $\mathbf{H}$.  However, unlike the edm, these voltage reversals are symmetric between the two electrodes.  Memory of these induction voltages through the charging of capacitors in the detector produces offset voltages that can obscure the edm voltage.  The voltage offset produced by this memory goes as the temporal area of the pulse (\mbox{V-s}) divided by the relevant RC time (s).  The offset is effectively reduced in the AC coupling of the amplifier by increasing R during the induction pulses.  A similar effect on the JFET input is reduced to an acceptable level by keeping the detector input impedance high ($10^{13}$~$\Omega$).

\begin{figure}
\includegraphics{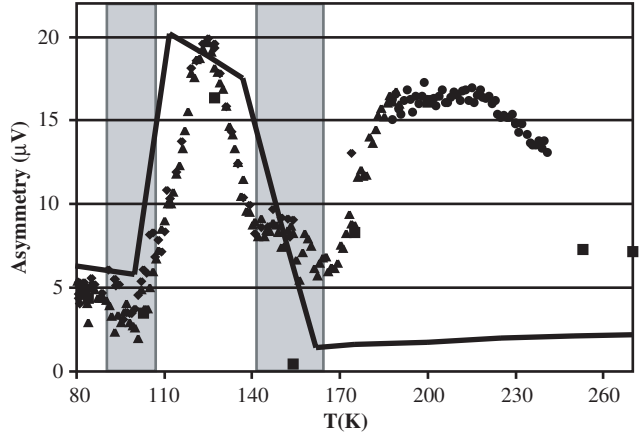}
\caption{Observed asymmetry in the difference voltage (D) between the two electrodes as a function of temperature with an applied ``square wave'' magnetic field of magnitude 53~Oe. The small diamonds, triangles and dots are data collected on three different days while scanning the temperature. The large squares are data collected with the sample well equilibrated at fixed temperatures. The solid line indicates the expected shape of the edm signal as a function of temperature. The gray rectangles denote regions where the temperature is too close to the compensation temperature of one of the ``C''s to be reliable.  In these regions the magnetic moment of one of the ``C''s is too small and domain creep follows the reversal of the applied field.\label{signalvtemp}}
\end{figure}

We construct a voltage ``asymmetry'' by taking the difference between the observed voltages for opposing circulations of the applied $\mathbf{H}$.  We construct ``asymmetries'' for each electrode ($A$ and $B$) and for the difference ($D = A-B$).  The edm voltage should manifest as an asymmetry in $D$.  A plot of the difference asymmetry as a function of temperature is shown in figure \ref{signalvtemp}.  These data are acquired in ``square wave'' mode with applied fields of 53~Oe.  Also shown is the expected edm signal based on the anticipated orientation of the Gd ions at various temperatures.  A large effect is seen which mimics the edm signature at low temperatures.  However, unlike the expected edm signal, this asymmetry remains large at temperatures above 180~K.  We believe that this asymmetry is produced by the imperfect reversal of a much larger quadratic magnetic field effect.

\begin{figure}
\includegraphics{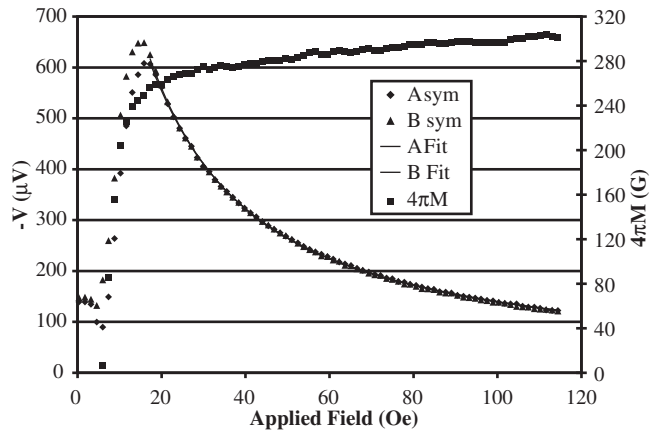}
\caption{The negative of the symmetric part of the electrode voltages ($A$ and $B$) and the magnetization as a function of $H$ with the application of a triangle wave.  Only the part of the cycle with increasing magnitude of $H$ is shown.  The solid lines are the asymptotic fit of the high field part of the electrode voltages to the function $V = a/H + b/H^2 + c$. Note that the voltage is similar for the two electrodes and has an onset that coincides with the rising of the magnetization.\label{symvapp}}
\end{figure}

To investigate this, we monitor the electrode voltages and the sample magnetization as a function of the applied field H, which is varied using a triangular wave.    We construct the symmetric and anti-symmetric parts of these signals by adding or subtracting the signals that are one half of a period out of phase.  Transient induction pulses due to $dB/dt$ appear clearly in the anti-symmetric parts of all of the signals.  The anti-symmetric part of the voltages on the electrodes are well correlated in time and magnitude with the current induced in a pick-up coil wrapped around the sample on the outside of the Faraday cage.  The symmetric parts of the electrode signals are shown in figure \ref{symvapp}.  The symmetric voltage changes are large.  They rise rapidly with the reversal of the sample magnetization and then fall off roughly as $H^{-1}$ at higher fields.  We refer to this effect as the ``$M$ even'' effect.  Unlike the anti-symmetric parts of the signals, this curve is not appreciably modified by changing the frequency of the triangle wave, further indicating that the symmetric part of the voltage is not induced by $dB/dt$.  

As a function of temperature, the difference asymmetries are significantly correlated with the amplitudes of the $M$ even voltages.  This suggests that the difference asymmetries may be due to imperfect reversal (perhaps due to local pinning of magnetic domains) of the much larger $M$ even effect. Higher fields suppress the size of the $M$ even voltage (Fig.~\ref{symvapp}) and may also result in an improved reversal of the effect due to suppression of domain pinning.  To achieve higher fields without excessive heating we switch to the ``pulsed'' mode of data acquisition. The observed difference asymmetry appears to decrease to some asymptotic value as the field strength increases.

We use the Marquardt-Levenberg algorithm to fit the high field pulsed difference asymmetries to the asymptotic functional form $V = b/H + V_0$, where $b$ and $V_0$ are constants.  Only values of $H$ larger than 125~Oe are included in the fit.  This high-field asymptotic form is suggested by studies of the approach to saturation of the magnetic domains in polycrystalline YIG~\cite{lefloch90} and selected because of the high quality of the resulting fit. Several data scans are taken at each temperature. The values of $b$ vary between about $-1.3$ and $-1.9$~\mbox{mV-Oe}, depending on the temperature. The averaged offsets ($V_0$) obtained from the fits are $V_0^{88K} = 1.18\pm0.11$~\mbox{$\mu$V}, $V_0^{127K} = 1.22\pm0.09$~\mbox{$\mu$V}, and $V_0^{178K} = 0.96\pm0.38$~\mbox{$\mu$V}. The large uncertainty at 178~K is due to non-statistical variations in $V_0$ with the rotation of the sample relative to the detector by $180^{\circ}$.   We extract the ``edm voltage'' by forming the difference $V_{edm} = V_0^{127K} - (V_0^{88K} + V_0^{178K})/2 = 0.14\pm0.22$~\mbox{$\mu$V}. We note that 0.1~\mbox{$\mu$V} on one of our electrodes corresponds to a charge of about 8~electrons. Application of the theory of Ref.~\onlinecite{mukhamedjanov03} to our sample allows us to infer from $V_{edm}$ a measurement of the electron edm of $(2\pm3)\times10^{-24}$~\mbox{e-cm}.

Systematic effects associated with capacitive pickup, external magnetic fields, circuit memory of the induction pulses, and domain creep have all been considered and are found to be negligible at our present level of sensitivity.  The result is limited by our ability to reliably extract the asymptotic values of $V_0$.  The large size of the $M$ even effect and the sensitivity of $V_0$ to the experimental configuration and chosen asymptotic form make this particularly difficult.  Nonetheless, our result does establish an upper bound on the electron edm, $d_e < 5\times10^{-24}$~\mbox{e-cm}, a factor of forty lower than the previous best value obtained from a solid-state system.  However, this bound remains more than 3 orders of magnitude larger than the best atomic limit.

Further improvement upon this limit in a solid will require dramatic suppression of the $M$ even effect.  We are presently engaged in an investigation of the source of this effect.  While a voltage with this symmetry can be produced by differential magnetostriction, the observed effect is more than an order of magnitude larger than theoretical upper bounds on its size~\cite{sushkov}. In addition, unlike magnetostriction, the observed effect does not have an extremum at zero magnetization, but appears to peak during the rotation mode (mode~III) of the magnetization process (Ref.~\onlinecite{mercier74}).  This suggests that the source of the voltage may be related to the relative angle between the net magnetization and the easy axis of the various crystallites. The functional dependence of the $M$ even effect is consistent with a voltage proportional to $M(M_{sat}-M)$, where the voltage is referenced to zero at $M_{sat}$, the saturation magnetization. Indeed, in the high field limit this expression approaches $M_{sat}^2a/H$, where $a$ is the magnetic hardness of the sample (Ref.~\onlinecite{lefloch90}).  This $H^{-1}$ dependence matches our observed high field behavior. We are presently investigating the dependence of the $M$ even effect on the polycrystalline nature of the sample, the toroidal geometry, and the relative proportions of Gd to Y.  Depending on the outcome of these investigations a single crystal sample, a change in geometry or doping, or an entirely different crystal (e.g.~LiGdF$_4$) may be necessary to adequately suppress this effect.

\begin{acknowledgments}
We wish to thank P. Grant, G. Gallo and R. Bartos for important technical support and Pacific Ceramics for the fabrication of our sample.  We thank D. DeMille, J. Friedman and O. Sushkov for important discussions. This work was supported by NSF grant 0244913, the Howard Hughes foundation, Amherst College and Los Alamos National Laboratory. 
\end{acknowledgments}

\bibliography{hunterpaper}

\end{document}